# Negative differential resistance and effect of defects and deformations in MoS$_2$ armchair nanoribbon MOSFET

Amretashis Sengupta*, Santanu Mahapatra

*Nano-Scale Device Research Laboratory, Dept. of Electronic Systems Engineering,*

*Indian Institute of Science, Bangalore – 560 012, India*

***Corresponding Author: E-mail:** amretashis@dese.iisc.ernet.in, **Fax:** +91-80-23600808*

***Abstract:*** *In this work we present a study on the negative differential resistance (NDR) behavior and the impact of various deformations (like ripples, twist, wrap) and defects like vacancies and edge roughness on the electronic properties of short-channel MoS$_2$ armchair nanoribbon MOSFETs. The effect of deformation (3-7$^o$ twist or wrap and 0.3-0.7 Å rippling amplitude) and defects on a 10 nm MoS2 ANR FET is evaluated by the density functional tight binding theory and the non-equilibrium Green`s function (DFTB-NEGF) approach. We study the channel density of states, transmission spectra and the I$_D$ – V$_D$ characteristics of such devices under the varying conditions, with focus on the negative differential resistance (NDR) behavior. Our results show significant change in the NDR peak to valley ratio (PVR) and the NDR window with such minor intrinsic deformations, especially with the rippling.*

**I. INTRODUCTION**

Following the demonstration of monolayer MoS$_2$ based MOSFET and logic [1,2] lot of interest has been generated on the computational study of such devices. [3]-[5] Such 2-D channel materials have shown to offer good MOSFET performance and considerable immunity to short channel effects such as drain induced barrier lowering (DIBL). An interesting phenomenon in such short channel MOSFET is the onset of negative differential resistance (NDR) with increasing quantum confinement, as the channel length get shorter and shorter. In case of nanoribbon (NR) MOSFET this NDR effect can be realized more easily owing to the lateral confinement in the NR structure. [6] In case of MoS$_2$ it is the armchair nanoribbon (ANR) which shows semiconducting behaviour, and for this reason it could be is useful for MOSFET applications. With such NDR effect in MoS$_2$ ANR MOSFET useful memory applications may be realized which could aid in further scaling of MOS memories.

However 2-D membranes like graphene, silicene, monolayer $MoS_2$ are reported to undergo a certain amount of minor deformations (like rippling/ buckling, crumpling etc). [7]-[9] These effects are understood to arise from various causes such as thermodynamic instabilities in such 2-D materials, the differences between thermal expansion coefficient of the film and substrate, and strains arising during transfer of 2-D films onto substrates. [7]-[9] In case of nanoribbons the most common types of deformations include the minor amount of twisting and wrapping. Though rather small in magnitude, such deformations could have a significant impact on the performance of $MoS_2$ ANR channel in terms of transmission properties and especially on the NDR characteristics of $MoS_2$ FET. Thus it could be worthwhile to investigate the NDR behaviour of MoS2 ANR FETs in the event of such defects and deformations.

In this work we present a density functional tight binding theory and the non-equilibrium Green`s function (DFTB-NEGF) study on the effect of various deformations (like rippling/periodic buckling, wrapping, twisting) on the performance and especially the NDR effect in a 10 nm channel length $MoS_2$ ANR MOSFET. The width of the ANR was considered to be 3.5nm. We further investigate the effect of minor variations in the ripple amplitude and angle of twist and wrap on the I-V characteristics and especially the associated NDR parameters like the peak current, peak-to-valley ratio (PVR), and the NDR window.

## II. METHODOLOGY

We consider monolayer $MoS_2$ armchair nanoribbon (ANR) as our channel material, with 10 nm channel length ($L_{Ch}$) and 3.5nm width. The ANR MOSFET is considered having a back gated configuration, with 3.5nm thick $HfO_2$ gate dielectric. The device schematic (not to scale) is given in Fig. 1(a). We consider doped $n^+$ regions as the source and the drain for our MOSFET with source/drain Fermi levels which are in alignment with the conduction band edge For simulating the conditions of intrinsic deformations of the ANR, we apply a very small degree of mechanical deformation (3-7$^o$ twist or wrap with transport direction as the wrapping or twisting axis and 0.3-0.7 Å ripple amplitude) on the $MoS_2$ ANR channel. The axis of the twist and the wrap lies along the channel of the MOSFET, while the ripple has a periodicity of 2 in the transport direction of the $MoS_2$ ANR. For defects we considered the edge roughness which is the most common imperfection in a nanoribbon structure, and random vacancies of Mo or S atoms in the ANR body. . In Fig. 1(b) we show the various defects and deformations considered in the $MoS_2$ ANR channel in our study. In Fig. 1(b) the magnitude of some of the deformations has been amplified for better visual representation.

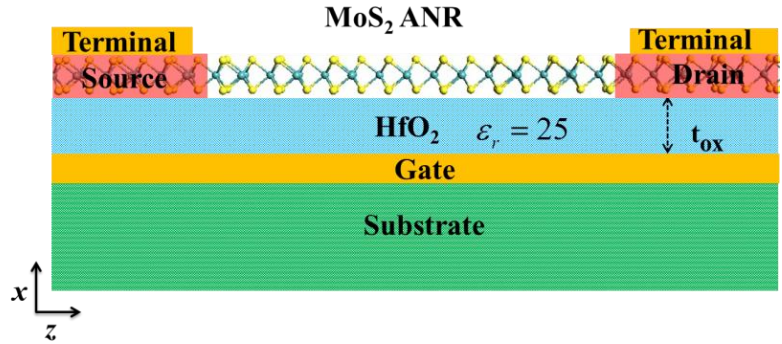

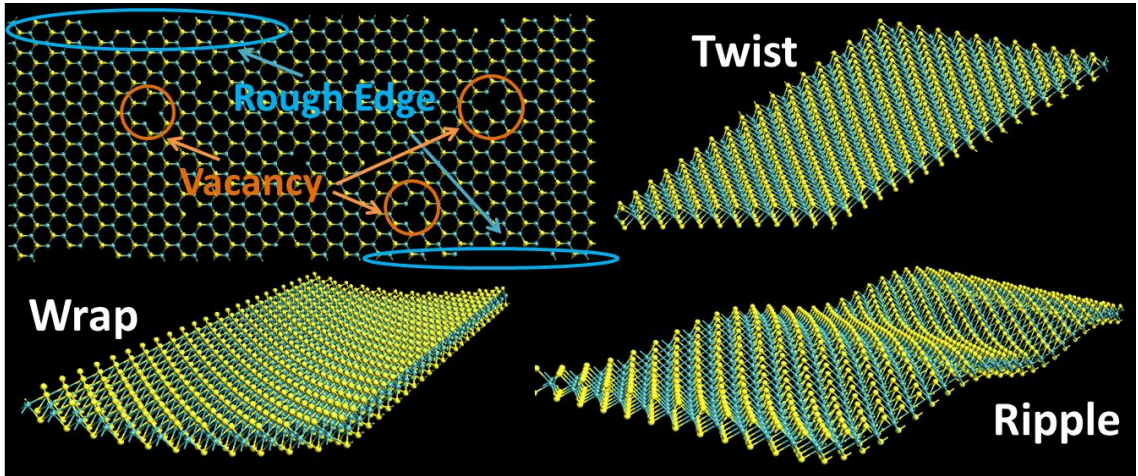

**Fig. 1. (a)** Device schematic (not to scale) of the back gated $MoS_2$ ANR MOSFET **(b)** schematic of various defects and deformations in $MoS_2$ ANR channel (magnitudes of twist, wrap and ripple exaggerated for better visibility)

This device configuration is simulated in Quantum Wise ATK using the self-consistent density functional tight-binding (DFTB) -NEGF method. This DFTB-NEGF method was chosen as it has emerged as a fast and computationally efficient method to simulate electronic transport in $MoS_2$ nanostructures like nanoribbons and tubes, which contain a large number of atoms. [10]-[15] For the DFTB study we use a 1x1x50 Monkhorst-Pack k-grid [16] for the device supercell and employ the semi-emperical Slater-Koster CP2K basis set for Mo-S in ATK. [14,15] The I-V characteristics for the different deformed conditions are evaluated using the NEGF formalism. [15]

In the NEGF method we proceed to solve the-Poisson Schrödinger equation of the system (described in Fig. 1) self-consistently. Setting up the self-energy matrices $\Sigma_S$ and $\Sigma_D$ for the source and drain contacts, the Green's function $G$ is constructed as

$$G(E) = [EI - H - \Sigma_S - \Sigma_D]^{-1} \qquad (1)$$

In (1) $I$ is the identity matrix. It is notable that here we consider the transport to be fully ballistic and therefore no scattering energy matrix has been incorporated into the Green's function. [15,17] From (1) parameters like the broadening matrices $\Gamma_S$ and $\Gamma_D$ and the spectral densities $A_S$ and $A_D$ are evaluated using the following relations

$$\Gamma_{S,D} = i\left[\Sigma_{S,D} - \Sigma^\dagger_{S,D}\right] \qquad (2)$$

$$A_{S,D}(E) = G(E)\Gamma_{S,D}G^\dagger(E) \qquad (3)$$

The density matrix $[\Re]$ used to solve the Poisson equation is given by

$$[\Re] = \int_{-\infty}^{\infty} \frac{dE}{2\pi}[A(E_{k,x})]f_0(E_{k,x} - \eta) \qquad (4)$$

where $A(E_{k,x})$ is the spectral density matrix $E_{k,x}$ the energy of the conducting level, and $\eta$ being the chemical potential of the contacts, $f_0$ is the Fermi function.

For NEGF in ATK, a multi-grid poisson solver is employed using Dirichlet boundary conditions on the left and right faces (i.e. the electrodes) and Neumann boundary conditions on the other faces (i.e. ANR edges) of the device. The electrode temperatures are considered 300 K. The carrier densities evaluated from the NEGF formalism are put into the Poisson solver to evaluate a more accurate guess of the self-consistent potential $U_{SCF}$ and the same is used to evaluate a more accurate value of number density of carriers $n_{tot}$. The converged values are used to evaluate the transmission matrix $\Im(E,V)$ as

$$\Im(E,V) = trace\,[A_S\Gamma_D] = trace\,[A_D\Gamma_S] \qquad (5)$$

For the transmission spectra we use the Krylov self-energy calculator [15] with the average Fermi level being set as the energy zero parameter. From this the ballistic drain current is easily evaluated as [14, 17, 18]

$$I_D = \left(\frac{4e}{h}\right) \int_{-\infty}^{+\infty} \Im(E,V) \left[ f_S\left(E_{k,x} - \eta_S\right) - f_D\left(E_{k,x} - \eta_D\right) \right] dE \quad (6)$$

In (6), $e$ is the electronic charge, $h$ is the Planck's constant, $f_S$ and $f_D$ are the Fermi functions in the source and drain contacts. $\eta_S$ and $\eta_D$ are the source and drain chemical potentials respectively. The factor 4 originates from the spin degeneracy and valley degeneracy in MoS$_2$ ANR. Equation (6) represents fully ballistic transport in the ANR FETs which holds good for short channel lengths below 15 nm. [17, 18]

## III. RESULTS & DISCUSSIONS

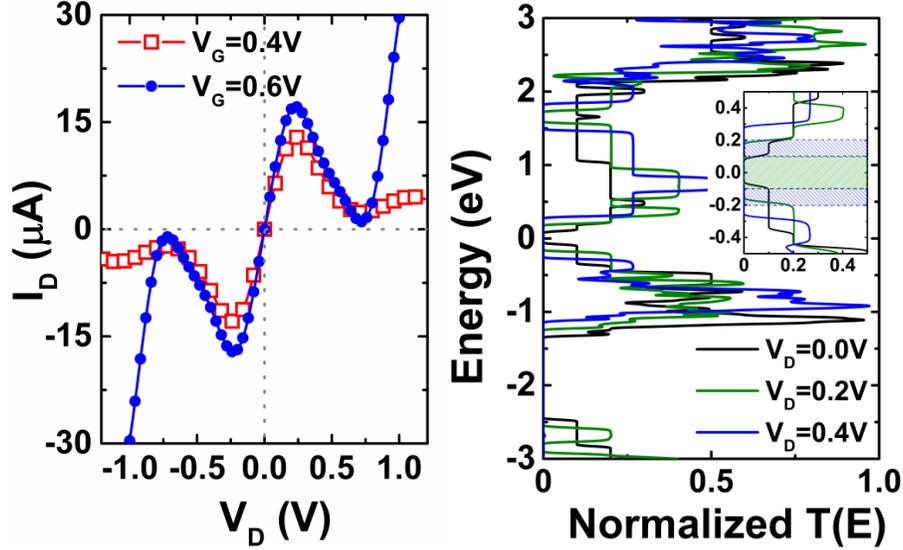

**Fig. 2 (a)** The NDR characteristics observed in 10nm MoS$_2$ ANR FET **(b)** The transmission spectra of a perfectly relaxed 10 nm MoS$_2$ ANR FET for a gate bias of 0.6 V and varying drain biases (Fermi level is set at zero), inset shows the bias windows for V$_D$ = 0.2, 0.4 V.

In Fig. 2(a) we see the $I_D - V_D$ characteristics of the MoS2 ANR MOSFET under fully relaxed condition (no deformation applied). The curve clearly displays the NDR characteristics of the device under a drain bias sweep from +1.2 to -1.2 V, with varying gate bias of 0.4 and 0.6 V. For an applied gate bias of 0.6 V, the 1$^{st}$ current peak lies at V$_D$ = 0.23 V and the value of this current is 17.16 µA. The 1$^{st}$ valley lies at 0.71 V with a current value of 1.06 µA. This gives a peak to valley ratio (PVR) of 16.23 and an NDR window of 0.48 V. For an applied gate voltage of 0.4 V this PVR is 5.42 and the window remains the same. The NDR behavior is almost symmetric for negative bias as well. The origin of the observed NDR characteristics can be understood from the transmission spectra of the device configuration which is shown in Fig. 2(b). Here it is observed that for a fixed gate bias of 0.6 V, as a small drain bias is applied more number of transmission modes appear near the Fermi level of the channel, which indicates a greater transmission transparency of the channel. As the drain voltage is increased further, the some of the transmission modes near the Fermi level seem to be suppressed. This leads to a drop in the transmission transparency of the channel, thus indicating NDR behavior. As the Fermi level ($E_F$) is set as zero in our calculations, the bias window (dashed regions in the inset of Fig. 2(b)) is effectively between $-V_D/2$ to $+V_D/2$. In Fig. 2(b) we also see that the area under the transmission spectra in the corresponding bias window initially increases from that of the zero drain bias value, but as the bias is increased further, the same area seems to decrease with transmission modes becoming more sharply resolved along particular energy levels, thus indicating a drop in current density.

As the MoS$_2$ ANR FET is subject to small amounts of elastic deformation like 5$^o$ twist, 5$^o$ wrap or 0.5 Å amplitude periodic ripples, the $I_D - V_D$ characteristics get affected by it as shown in Fig. 3. Here we see that with the application of deformation the PVR and the NDR window of the MOSFET are significantly influenced by various defects and deformations. The most significant changes are seen for the vacancy, rough edge and rippling in particular. While the reduction in PVR for twist or wrap lies in the range of 60% for rippling the change is about 90% from its original (relaxed) value. For defect structures like edge roughness and vacancy, this change is between 75-85%. For the NDR window, the twist and wrap do not have an impact as significant as ripple which reduces the window from its original value by 54%.

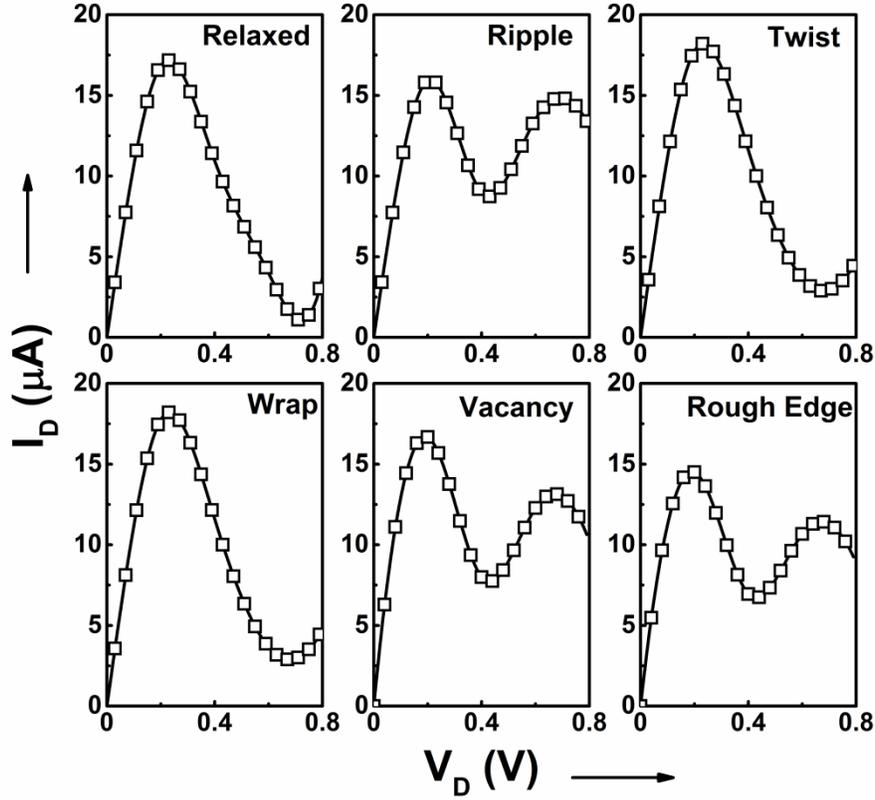

**Fig. 3.** The $I_D$-$V_D$ characteristics for $V_G = 0.6$ V observed in 10nm MoS$_2$ ANR FET under relaxed and elastically deformed or defect conditions.

This greater impact of ripples on the device characteristics is also evident from the corresponding DOS and the transmission spectra of the devices under the various types of elastic deformation shown in Fig. 4 (a) and (b). In Fig. 4 (a), we see that while the DOS of the ANR channel becomes more discretized with all the three kinds of deformations, it is also clear that the highest density of states in the conduction band and near the Fermi level occur for the rippled case. This causes a slight difference in the corresponding transmission spectra of the channel material, as shown in Fig. 4(b). Here among the various conditions, the rippled MOSFET shows the largest number of available transmission modes near the conduction band minima compared to the other devices.

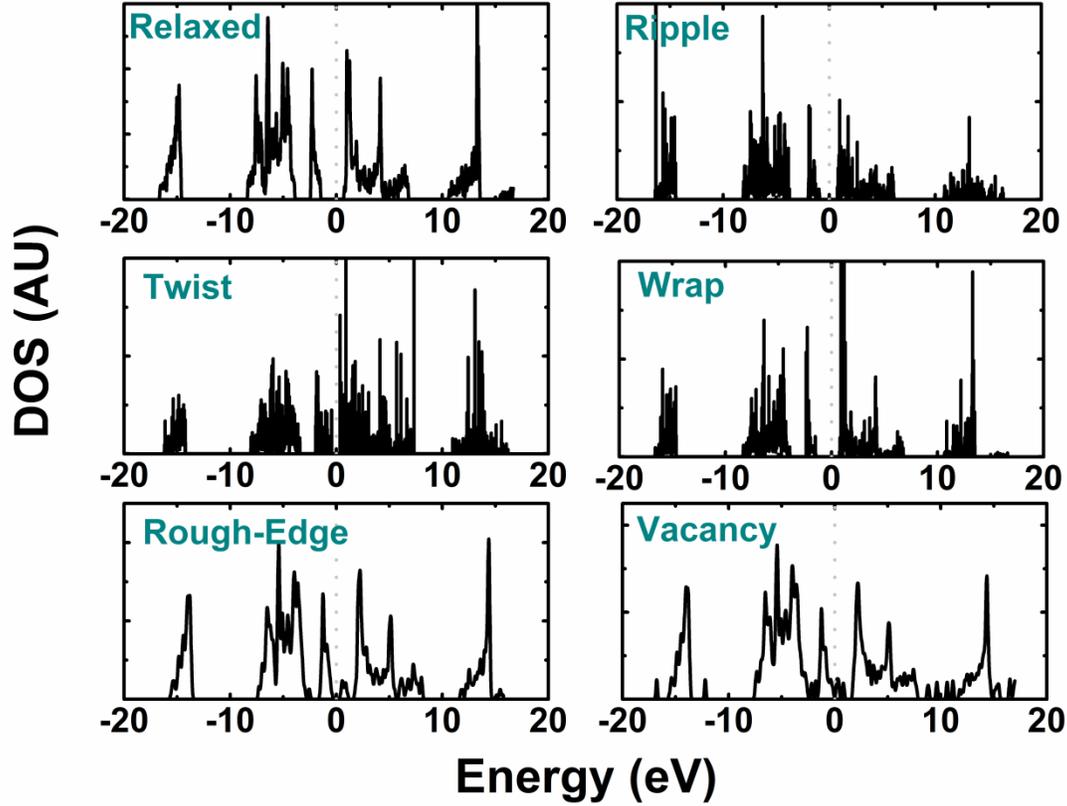

**Fig. 4. (a)** Density of states of the MoS$_2$ ANR channel under different conditions.

Even in the unbiased condition ($V_D = 0$ V) the rippled channel offers a larger number of transmission modes above the Fermi level (set at zero), than the twisted or the wrapped conditions. In case of the vacancies and the rough edges, the distribution of DOS somewhat more balanced in both the VB and the CB sides near the Fermi level, though comparatively quite higher number of states are present near the band edges in both defects as compared to the relaxed case. The vacancies and the rough edge ANR channels both have quite similar kinds of DOS and transmission spectra.

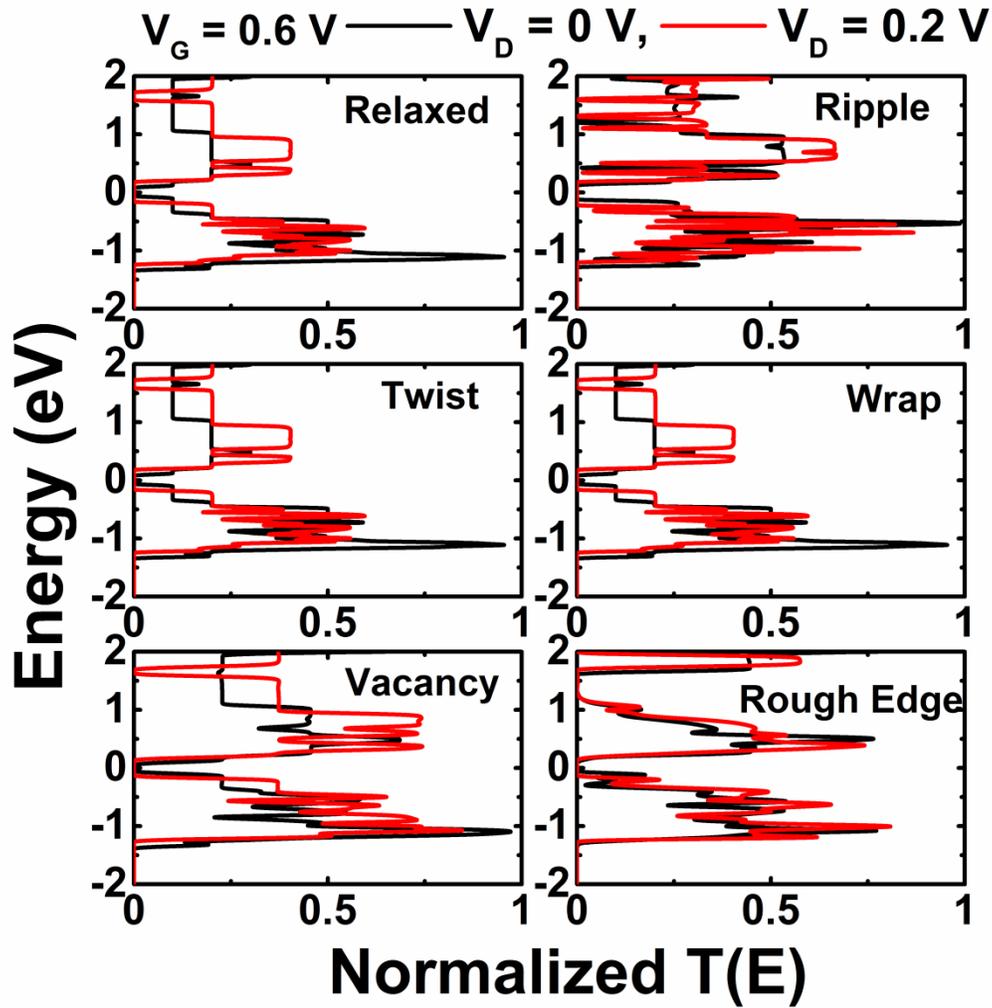

**Fig. 4. (b)** Transmission spectra of the MOSFET under the various deformed conditions (Fermi level is set at zero).

We now come to the impact of the variation of the NDR parameters with varying degree of deformation. In Fig.5(a)-(c), we show the variations of these characteristics with varying angle of twist and wrap $3 - 7^o$ and the ripple amplitude from $0.3 - 0.7$ Å.

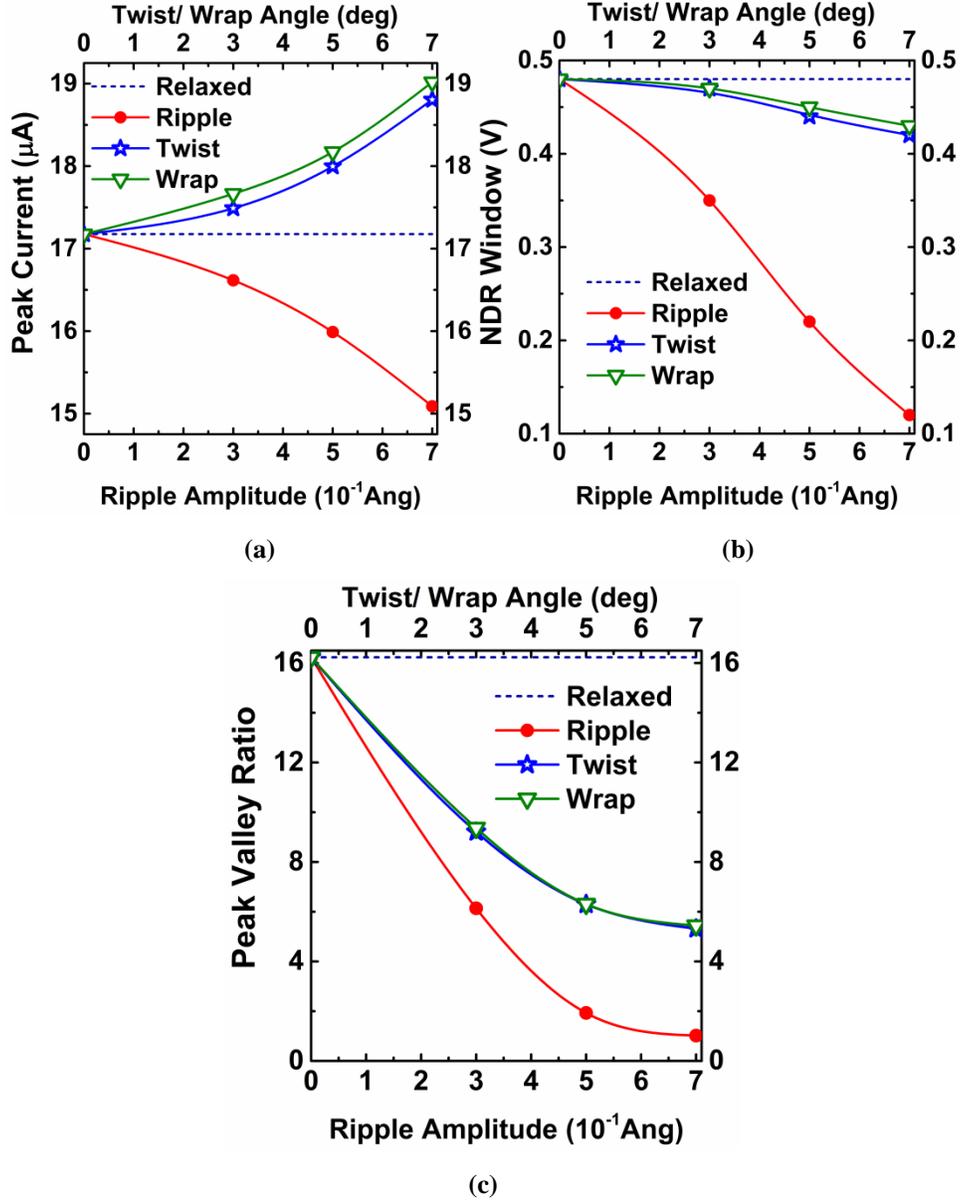

**Fig. 5.** The variation of **(a)** 1st Peak current **(b)** NDR window and **(c)** the Peak-to-valley ratio (PVR) with varying magnitude of applied twist, wrap and rippling.

Here we see that with the increasing magnitude of twist and wrap, there is an increase in the current corresponding to the 1st NDR peak, while for increasing ripple amplitude there is a decrease in the same. In all the cases the NDR window is reduced, however the magnitude of reduction for the rippled case is far greater than the other deformations. Also a significant reduction of the PVR is observed for all the three cases but it is only in the case of ripples that the PVR could come down to 1.04 (i.e. the NDR effect becomes significantly suppressed). This observation is of considerable significance as the change in the NDR behaviour in MoS$_2$ ANR MOSFETs owing to mechanical deformations could also lead to development of sensors based on such devices.

## IV. CONCLUSION

In summary, the impact of elastic deformation in terms of varying twist, wrap and ripple and defects like edge roughness and vacancies on the I-V characteristics in ballistic short channel MoS$_2$ ANR MOSFET was studied. The transport in such short channel MoS$_2$ ANR FET was investigated using the DFTB-NEGF formalism. In all cases sizeable NDR effect was observed in the devices. Significant changes in NDR parameters such as PVR and NDR window was observed with elastic deformation. Among such small deformations, the ripples and the defects seemed to affect the NDR behavior more profoundly than the others. It was observed that with 0.7 Å amplitude of rippling with periodicity of 2 the NDR behavior effectively vanishes for 10nm MoS$_2$ ANR MOSFETs. The NDR behaviour in short channel MoS$_2$ ANR MOSFET when experimentally realized could be highly useful in the miniaturization of various circuits such as memory cells, amplifiers and RF oscillators. Moreover, the fact that small magnitudes of ripples, wrap, twist can induce significant change in the NDR behaviour in MoS2 MOSFETs could also lead to mechanical sensors based on such devices.

**ACKNOWLEDGEMENT**

A.S. thanks Department of Science and Technology (DST), Govt. of India, for his DST Post-doctoral Fellowship in Nano Science and Technology. The work was supported by DST, Govt. of India, under grant no. SR/S3/EECE/0151/2012.

**FIGURE CAPTIONS**

**Fig. 1. (a)** Device schematic (not to scale) of the back gated MoS$_2$ ANR MOSFET **(b)** schematic of various defects and deformations in MoS$_2$ ANR channel (magnitudes of twist, wrap and ripple exaggerated for better visibility)

**Fig. 2 (a)** The NDR characteristics observed in 10nm MoS$_2$ ANR FET **(b)** The transmission spectra of a perfectly relaxed 10 nm MoS$_2$ ANR FET for a gate bias of 0.6 V and varying drain biases (Fermi level is set at zero), inset shows the bias windows for V$_D$ = 0.2, 0.4 V .

**Fig. 3.** The I$_D$ -V$_D$ characteristics for V$_G$ = 0.6 V observed in 10nm MoS$_2$ ANR FET under relaxed and elastically deformed or defect conditions.

**Fig. 4. (a)** Density of states of the MoS$_2$ ANR channel under different conditions. **(b)** Transmission spectra of the MOSFET under the various deformed conditions (Fermi level is set at zero).

**Fig. 5.** The variation of **(a)** 1$^{st}$ Peak current **(b)** NDR window and **(c)** the Peak-to-valley ratio (PVR) with varying magnitude of applied twist, wrap and rippling.